\documentclass[12pt]{iopart}
\usepackage{iopams}
\usepackage{graphics}
\usepackage[next]{inputenc}
\usepackage[dvips]{epsfig}
\begin{document}
\title{Transport of charge and spin in the weak link between misoriented $PrOs_4Sb_{12}$ superconductors}
\author{G. Rashedi$^{1}$\footnote[7]{rashedy@iasbs.ac.ir} and Yu. A. Kolesnichenko$^{1,2}$}
\address{$^1$ Institute for Advanced Studies in Basic Sciences, 45195-159,
Zanjan, Iran\\ $^2$ B.Verkin Institute for Low Temperature Physics
 Engineering of National Academy of Sciences of Ukraine, 47,
  Lenin ave , 61103, Kharkov, Ukraine}
\date{\today}
\begin{abstract}
Recently, the ``$(p+h)-$wave'' form of pairing symmetry has been
proposed for the superconductivity in $PrOs_{4}Sb_{12}$ compound
[Parker D, Maki K and Haas S, \textbf{cond-mat/0407254}]. In the
present paper, a stationary Josephson junction as a weak-link
between $PrOs_{4}Sb_{12}$ triplet superconductors is theoretically
investigated. The quasiclassical Eilenberger equations are
analytically solved. The spin and charge current-phase diagrams
are plotted and the effect of misorientation between crystals on
the spin current, and spontaneous and Josephson currents is
studied. It is found that such experimental investigations of the
current-phase diagrams can be used to test the pairing symmetry in
the above-mentioned superconductors. It is shown that this
apparatus can be applied as a polarizer for the spin current.
\end{abstract}
\maketitle
\section{Introduction}
The pairing symmetry of the recently discovered superconductor
compound $PrOs_{4}Sb_{12}$ is an interesting topic of research in
the field of superconductivity \cite{Maki1,Maki2,Maki3}.
Superconductivity in this compound was discovered in papers
\cite{Bauer1,Bauer2,Kotegawa} and two different phases ($A$ and
$B$) have been considered for this kind of superconductor in
Refs.\cite{Maki2,Nakajima}. Although authors of \cite{Maki2} at
first considered the spin-singlet ``$(s+g)-$wave'' pairing
symmetry for this superconductor, later it was specified that the
spin-triplet is the real pairing symmetry of the $PrOs_{4}Sb_{12}$
complex \cite{Maki1,Tou}. Using the Knight shift in NMR
measurement authors of paper \cite{Tou}, estimated the
spin-triplet pairing symmetry for the superconductivity in $PrOs
_{4}Sb_{12}$. Consequently, the ``$(p+h)-$wave'' model of the
order parameter was proposed for the pairing symmetry of the
superconductivity in
$PrOs_{4}Sb_{12}$ compound, recently \cite{Maki1}. In the paper \cite{Maki1}%
, the self-consistent equation for the superconducting gap $\Delta
\left( T\right) $ (BCS gap equation) has been solved for the
finite temperature $T$ numerically and for the temperatures $T$
close to zero and the critical
temperature $T_{c}$ analytically. For this compound, using the ``$(p+h)-$%
wave'' symmetry for the order parameter vector (gap function), the
value of the $\Delta \left( 0\right) $ has been obtained for both
$A$ and $B$-phases, in terms of the critical temperature. In
addition, the dependence of $\Delta \left( T\right) $ in the
temperature limit of $T\rightarrow 0$ and $T\rightarrow T_{c}$
have been obtained. Authors of paper \cite{Maki1} have
investigated the temperature dependence of critical field,
specific heat and heat conductivity. Also, the Josephson effect in
the point contact between triplet superconductors with $f-$wave
triplet pairing has been studied in Ref.\cite{Mahmoodi}. In this
paper the effect of misorientation on the charge transport has
been studied and a spontaneous current tangential to the interface
between the $f$-wave superconductors has been observed.
Additionally, the spin-current in the weak-link between the
$f$-wave superconductors has been investigated in our paper
\cite{Rashedi1}. In the paper \cite{Rashedi1}, this kind of
weak-link device has been proposed as the filter for polarization
of the spin-current. These weak-link structures have been used to
demonstrate the order parameter symmetry in Ref.
\cite{Stefanakis}.\newline In the present paper, the ballistic
Josephson weak-link via an interface between two bulks of
``$(p+h)-$wave'' superconductor with different orientations of the
crystallographic axes is investigated. It is shown that the spin
and charge current-phase diagrams are totally different
from the current-phase diagrams of the point-contacts between conventional ($%
s$-wave) superconductors \cite{Kulik}, high $T_{c}$ ($d$-wave)
superconductors \cite{Coury} and from the charge and spin-current
phase diagrams in the weak-link between the $f$-wave
superconductors \cite{Mahmoodi,Rashedi1}. We have found that in
the weak-link structure between the ``$(p+h)-$wave''
superconductors, the spontaneous current parallel to the
interface, as the characteristic of unconventional
superconductivity, can be present. The effect of misorientation on
the spontaneous, Josephson and spin currents for the different
models of the paring symmetry ($A-$ and $B-$ phases in
Fig.\ref{phases}) are investigated. It is possible to find the
value of the phase difference, at which the Josephson current is
zero, but the spontaneous current tangential to the interface is
present. In some configurations and at the zero phase difference,
the Josephson current is not zero but has a finite value. This
finite value corresponds to a spontaneous phase difference, which
is related to the misorientation between the gap vectors. Finally,
it is observed that at certain values of the phase difference
$\phi ,$ at which the charge current is zero, the spin current is
present and vise versa. In addition, in the configuration in which
both gap vectors are directed along the $\hat{\mathbf{c}}\perp
\hat{\mathbf{n}}$ axis ($\hat{\mathbf{n}}$ is the normal to the
interface unit vector), only the normal to the interface spin
current $\hat{\mathbf{s}}_{\hat{\mathbf{n}}}$ can be present and
the other terms of the spin current are absent ($\hat{\mathbf{s}}$
is the spin vector of electrons).
\begin{figure}[h]
\includegraphics[width=5cm]{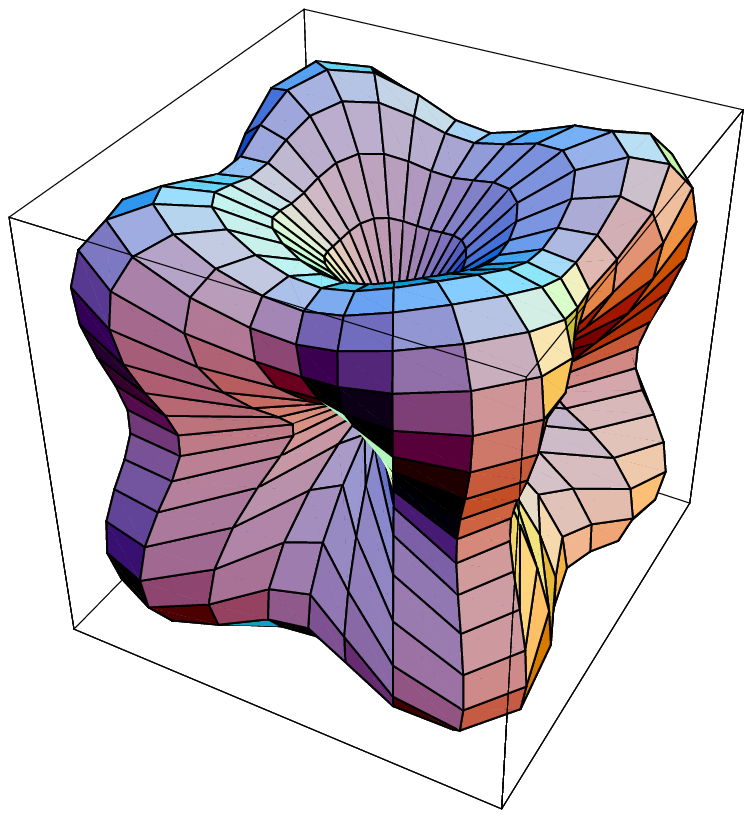}
\includegraphics[width=5cm]{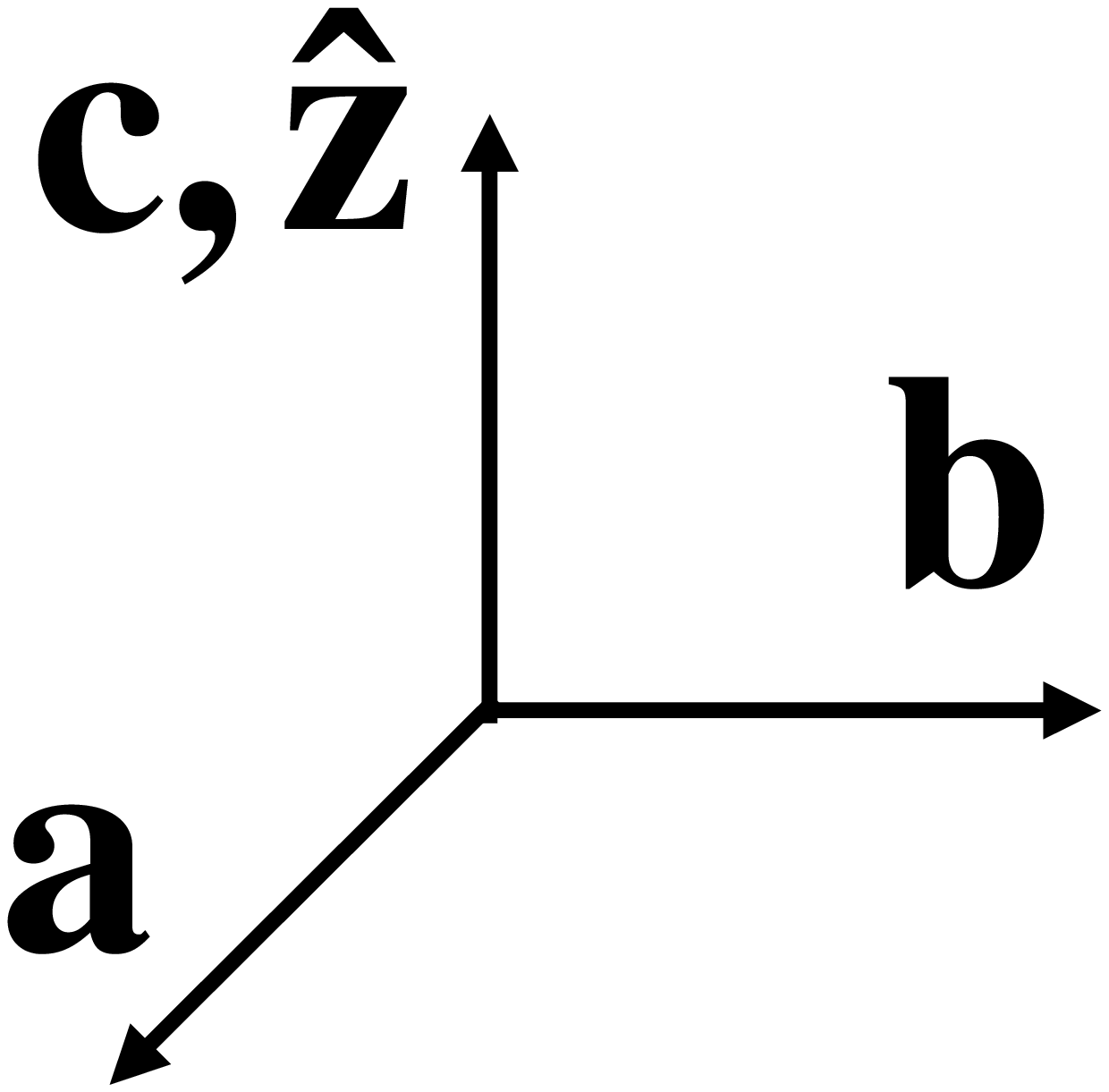}
\includegraphics[width=5cm]{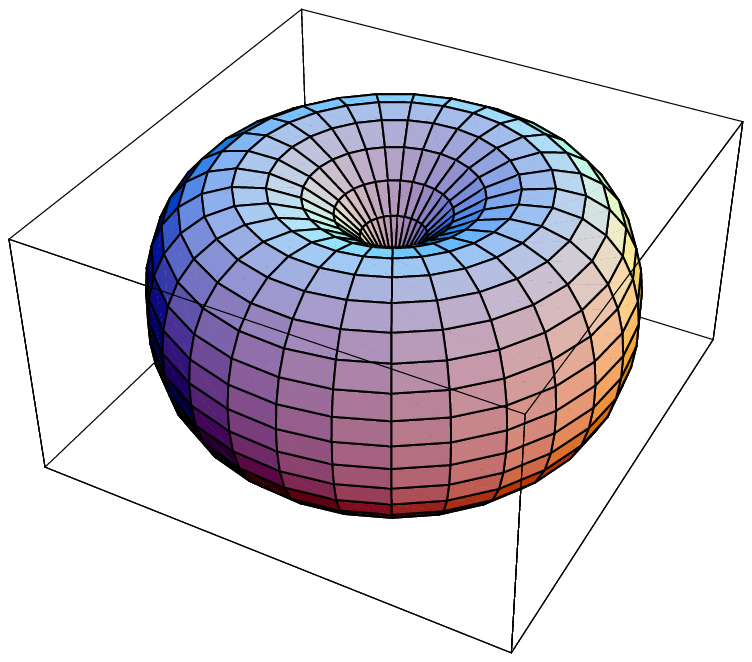}
\caption{$A-$phase (left), $B-$phase (right) order parameters and
 direction of $\mathbf{a}$, $\mathbf{b}$, $\mathbf{c}$ and
$\mathbf{\hat{z}}$ unit vectors (middle) \cite{Maki1}. $A-$ and
$B-$ phases are high field (high temperature) and low field (low
temperature) phases, respectively \cite{Goryo}.}
\end{figure}\label{phases}Consequently, this structure can be used as a filter for
polarization of the spin transport. Furthermore, our analytical
and numerical calculations have shown that the misorientation is
the origin of the spin current.\newline

The organization of the rest of this paper is as follows. In
Sec.\ref{section2} we describe our configuration, which has been
investigated. For a non-self-consistent model of the order
parameter, the quasiclassical Eilenberger equations
\cite{Eilenberger} are solved and suitable Green functions have
been obtained analytically. In Sec.\ref{section3} the obtained
formulas for the Green functions have been used for calculation
the charge and spin current densities at the interface. An
analysis of numerical results will be done in Sec.\ref{section4}.
The paper will be finished with some conclusions in
Sec.\ref{section5}.

\section{Formalism and Basic Equations}
\label{section2}
\begin{figure}[tbp]
\includegraphics[width=0.5\columnwidth]{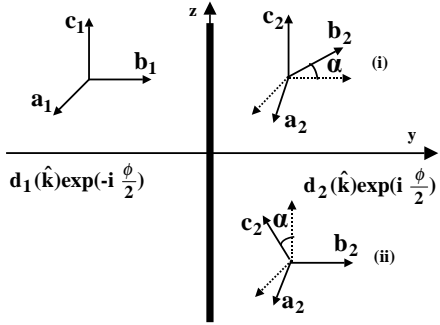}
\caption{Scheme of a flat interface between two superconducting
bulks, which are misoriented by angle $\protect\alpha $. In
geometry (i), the $ab$-plane on the right side is rotated as much
as $\alpha$ around the $c$-axis and in geometry (ii) the $c$-axis
on the right side is rotated around the $b$-axis.  $ab-$planes and
$c-$axis have been shown in Fig.\ref{phases}.} \label{fig1}
\end{figure}

We consider a model of a flat interface $y=0$ between two misoriented ``$%
(p+h)-$wave'' superconducting half-spaces (Fig.\ref{fig1}) as a
ballistic Josephson junction. In the quasiclassical ballistic
approach, in order to calculate the current, we use
``transport-like'' equations \cite{Eilenberger} for the
energy integrated Green matrix $\breve{g}\left( \mathbf{\hat{v}}_{F},%
\mathbf{r},\varepsilon _{m}\right) $
\begin{equation}
\mathbf{v}_{F}\nabla \breve{g}+\left[ \varepsilon _{m}\breve{\sigma}_{3}+i%
\breve{\Delta},\breve{g}\right] =0,  \label{Eilenberger}
\end{equation}
and the normalization condition $\breve{g}\breve{g}=\breve{1}$, where $%
\varepsilon _{m}=\pi T(2m+1)$ are discrete Matsubara energies $m=0,1,2...$, $%
T$ is the temperature, $\mathbf{v}_{F}$ is the Fermi velocity and $\breve{%
\sigma}_{3}=\hat{\sigma}_{3}\otimes \hat{I}$ in which $\hat{\sigma}%
_{j}\left( j=1,2,3\right) $ are Pauli matrices. Also the matrix
structure of the off-diagonal self energy $\breve{\Delta}$ in the
Nambu space is
\begin{equation}
\breve{\Delta}=\left(
\begin{array}{cc}
0 & \mathbf{d}\hat{\mathbf{\sigma }}i\hat{\sigma}_{2} \\
i\hat{\sigma}_{2}\mathbf{{d^{\ast }}\hat{\sigma}} & 0
\end{array}
\right),
\end{equation}
\label{order parameter} where, the gap vector $\mathbf{d}$, is the
three dimensional counterpart of the BCS energy-gap function
$\Delta$ for the case of triplet superconductivity.
The Green matrix $\breve{g}$ can be written in the form:
\begin{equation}
\breve{g}=\left(
\begin{array}{cc}
g_{1}+\mathbf{g}_{1}\mathbf{\hat{\sigma}} & \left( g_{2}+\mathbf{g}_{2}\hat{%
\mathbf{\sigma }}\right) i\hat{\sigma}_{2} \\
i\hat{\sigma}_{2}\left( g_{3}+\mathbf{g}_{3}\hat{\mathbf{\sigma
}}\right)
&i\hat{\sigma}_{2}(g_{1}-\mathbf{g}_{1}\hat{\mathbf{\sigma
}})i\hat{\sigma}_{2}
\end{array}
\right).
\end{equation}
\label{Green's function} Here,
$g_{1}+\mathbf{g}_{1}\mathbf{\hat{\sigma}}$ and
$(g_{2}+\mathbf{g}_{2}\hat{\mathbf{\sigma }})i\hat{\sigma}_{2}$
[$i\hat{\sigma}_{2}(g_{3}+\mathbf{g}_{3}\hat{\mathbf{\sigma }})$]
are normal and anomalous Green function matrix, respectively. The
terms of $g_i$ and $\mathbf{g}_{ij}$ are coefficients of Green
function matrix expansion in terms of the unit matrix and Pauli
$2\times 2$ matrices in the Nmabu space. Also, the terms $g_1$ and
$\mathbf{g_1}$ determine the charge and spin current densities
through the equations (\ref{charge-current}) and
(\ref{spin-current}), respectively and $\mathbf{g_2}$ and
$\mathbf{g_3}$ are used to determine the gap vector, $\mathbf{d}$,
using the self consistent relations (Eq.\ref{self-consistent} and
it's conjugate). It is remarkable that, in this paper, the unitary
states, for which $\mathbf{d\times d}^{\ast }=0,$ is investigated.
Also, the unitary states vectors $\mathbf{d}_{1,2}$ can be written
as
\begin{equation}
\mathbf{d}_{n}=\mathbf{\Delta }_{n}\exp i\psi _{n},
\end{equation}
where $\mathbf{\Delta }_{1,2}$ are the real vectors in the left
and right sides of the junction.\ The gap (order parameter) vector
$\mathbf{d}$ has to be determined from the self-consistency
equation:
\begin{equation}
\mathbf{d}\left( \mathbf{\hat{v}}_{F},\mathbf{r}\right) =2\pi
TN\left(
0\right) \sum_{m}\left\langle V\left( {\mathbf{\hat{v}}}_{F},{\mathbf{\hat{v}%
}}_{F}^{\prime }\right) \mathbf{g}_{2}\left(
{\mathbf{\hat{v}}}_{F}^{\prime },\mathbf{r},\varepsilon
_{m}\right) \right\rangle  \label{self-consistent}
\end{equation}
where $V\left(
{\mathbf{\hat{v}}}_{F},{\mathbf{\hat{v}}}_{F}^{\prime }\right) $,
is a potential of pairing interaction, $\left\langle
...\right\rangle $ stands for averaging over the directions of an
electron
momentum on the Fermi surface ${\mathbf{\hat{v}}}_{F}^{\prime }$ and $%
N\left( 0\right) $ is the electron density of states at the Fermi
level of energy. Solutions to Eqs.(\ref{Eilenberger}) and
(\ref{self-consistent}) must satisfy the conditions for Green
functions and vector $\mathbf{d}$ in the bulks of the
superconductors far from the interface as follow:
\begin{eqnarray}
\breve{g}\left( \pm \infty \right) &=&\frac{\varepsilon _{m}\breve{\sigma}%
_{3}+i\breve{\Delta}_{2,1}}{\sqrt{\varepsilon _{m}^{2}+\left|
\mathbf{\Delta
}_{2,1}\right| ^{2}}};  \label{Bulk solution} \\
\mathbf{d}\left( \pm \infty \right) &=&\mathbf{\Delta }_{2,1}\left( \mathbf{%
\hat{v}}_{F}\right) \exp \left( \mp \frac{i\phi }{2}+i\psi
_{2,1}\right) , \label{Bulk order parameter}
\end{eqnarray}
where $\phi $ is the external phase difference between the order
parameters of the bulks. Eqs. (\ref{Eilenberger}) and
(\ref{self-consistent}) have to be supplemented by the continuity
conditions at the interface between superconductors. For all
quasiparticle trajectories, the Green functions satisfy the
boundary conditions both in the right and left bulks as well as at
the interface.\newline The system of equations
(\ref{Eilenberger})and (\ref{self-consistent}) can be solved only
numerically.
\begin{figure}[tbp]
\includegraphics[width=0.5\columnwidth]{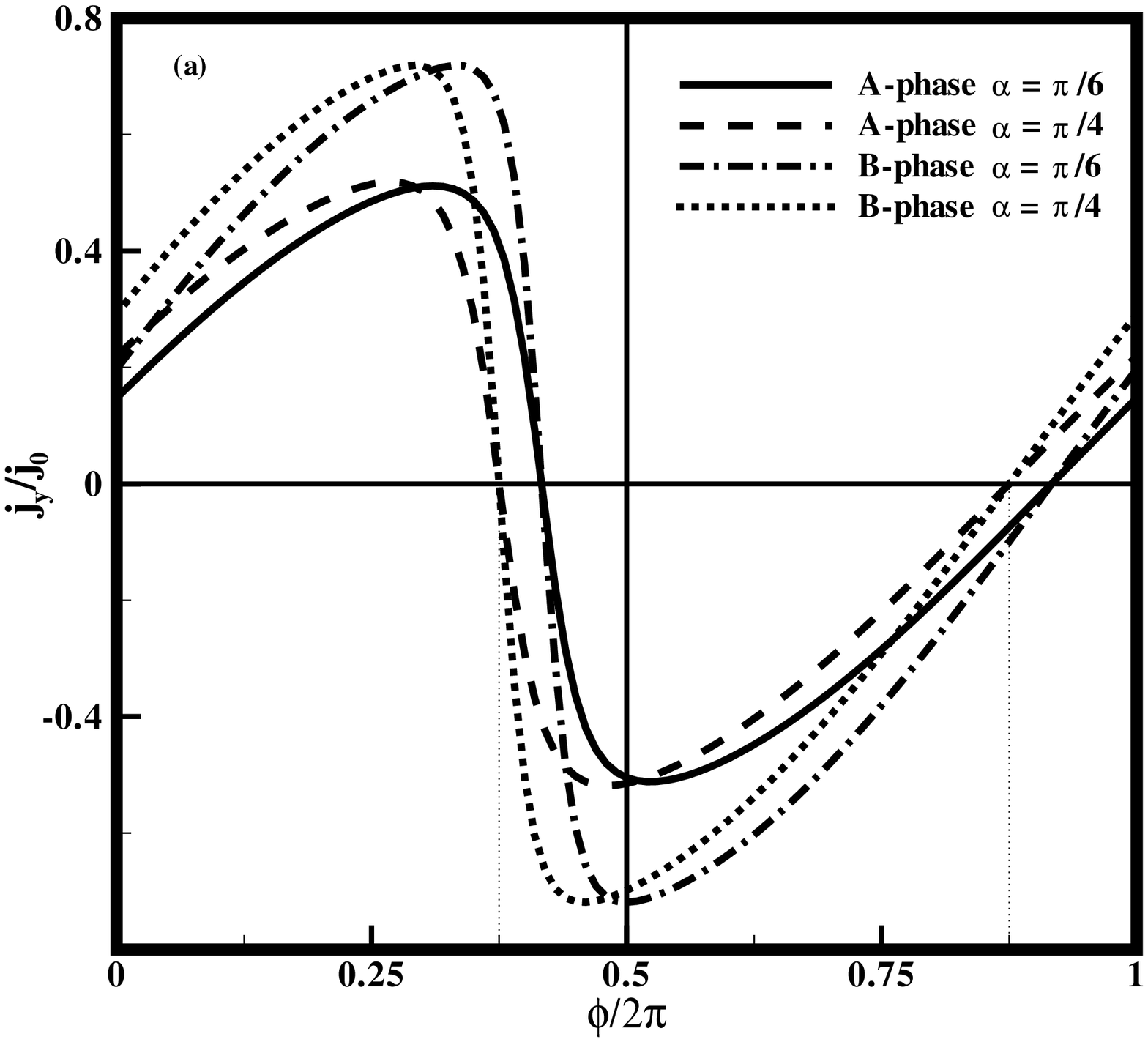}
\includegraphics[width=0.5\columnwidth]{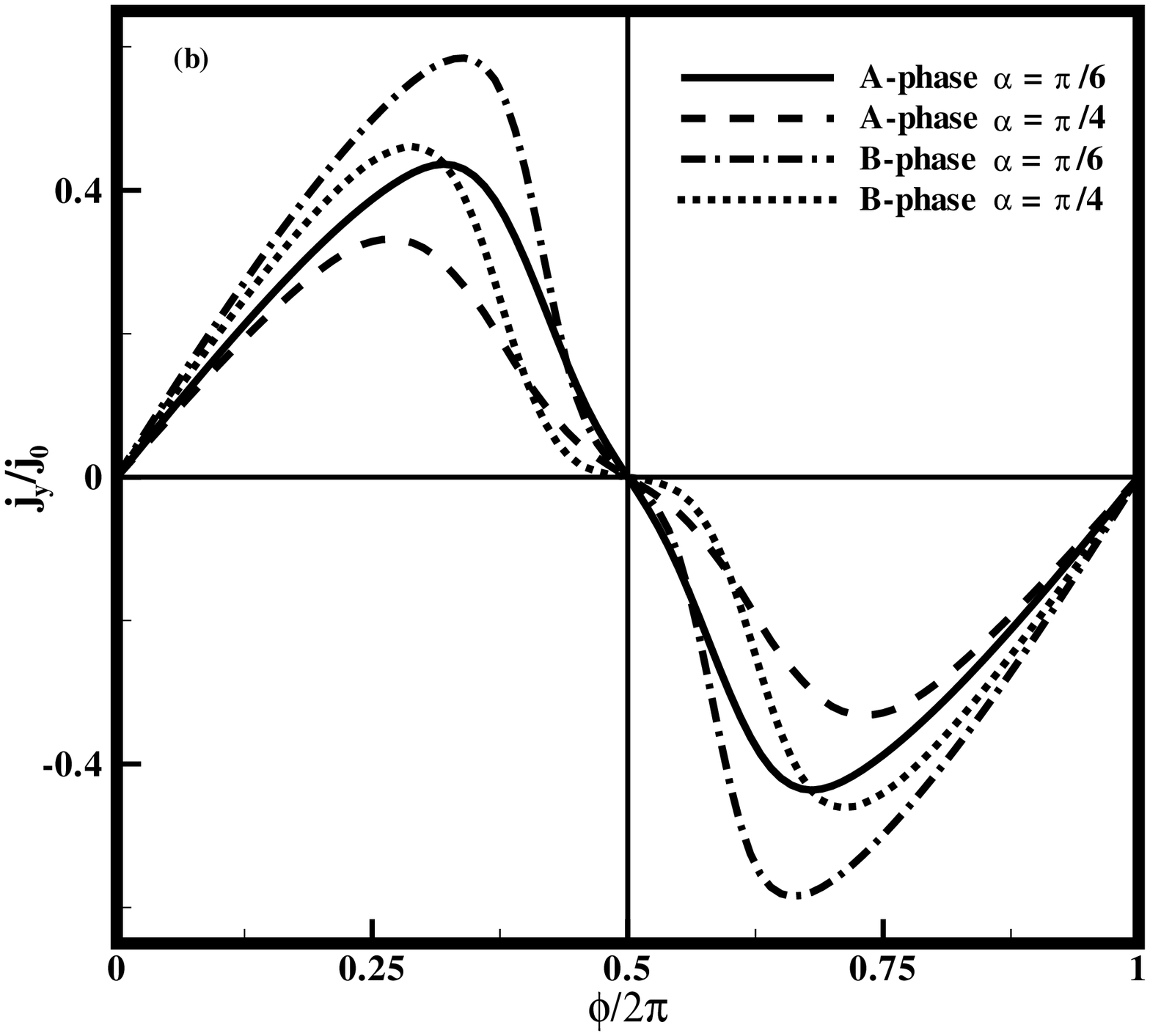}
\caption{Component of charge current normal to the interface
(Josephson current) versus the phase difference $\protect\phi $
for $A$ and $B-$phases, $T/T_{c}=0.08$ and different
misorientations. Part (a) is plotted for geometry (i) and part (b)
for geometry (ii). Charge currents are given in units of
$j_{0}=\frac{\protect\pi }{2}eN(0)v_{F}\Delta _{0}(0).$}
\label{fig2}
\end{figure}
For unconventional superconductors such solution
requires the information of the function $V\left( {\mathbf{\hat{v}}}_{F},{%
\mathbf{\hat{v}}}_{F}^{\prime }\right) $. This information, as
that of the nature of unconventional superconductivity in novel
compounds, in most cases is unknown. Usually, the spatial
variation of the gap vector and its dependence on the momentum
direction can be separated in the form of $\Delta
({\mathbf{\hat{v}}}_{F},y)=\Delta ({\mathbf{\hat{v}}}_{F})\Psi
(y)$. It has
been shown that the absolute value of a self-consistent order parameter and $%
\Psi (y)$ are suppressed near the interface and at the distances
of the order of the coherence length, while its dependence on the
direction in the momentum space ($\Delta
({\mathbf{\hat{v}}}_{F})$) remains unaltered \cite{Barash}.
Consequently, this suppression doesn't influence the Josephson
effect drastically. This suppression of the order parameter keeps
the current-phase dependence unchanged but, it changes the
amplitude value of the current. For example, it has been verified
in Refs.\cite{Coury} for the junction between unconventional
$d$-wave, in Ref.\cite{Barash} for the case of ``$f$-wave''
superconductors and in Ref.\cite{Viljas} for pinholes in $^{3}He$
that, there is a good qualitative agreement between
self-consistent and non-self-consistent results. Also, it has been
observed that the results of the non-self-consistent investigation
of $D-N-D$ structure in \cite{Faraii} are coincident with the
experimental results of \cite{Freamat} and the results of the
non-self-consistent model in \cite{Yip} are similar to the
experiment \cite{Backhaus}. Consequently, despite the fact that
self-consistent numerical results cannot be applied directly for a
quantitative analysis of the real experiment, only a qualitative
comparison of calculated and experimental current-phase relations
is possible. In our calculations, a simple model of the constant
order parameter up to the interface is considered and the pair
breaking and the scattering on the interface are ignored. We
believe that under these strong assumptions our results describe
the real situation qualitatively. In the framework of such model,
the analytical expressions for the current can be obtained for an
arbitrary form of the order parameter.
\section{Analytical results.}
\label{section3} The solution of Eqs.(\ref{Eilenberger}) and
(\ref{self-consistent}) allows us to calculate the charge and spin
current densities. The expression for the charge current is:
\begin{equation}
\mathbf{j}_{e}\left( \mathbf{r}\right) =2i\pi eTN\left( 0\right)
\sum_{m}\left\langle \mathbf{v}_{F}g_{1}\left( \mathbf{\hat{v}}_{F},\mathbf{r%
},\varepsilon _{m}\right) \right\rangle,  \label{charge-current}
\end{equation}
and for the spin current we have:
\begin{equation}
\mathbf{j}_{s_{i}}\left( \mathbf{r}\right) =2i\pi (\frac{\hbar
}{2})TN\left(
0\right) \sum_{m}\left\langle \mathbf{v}_{F}\left( \mathbf{{\hat{e}}}_{i}%
\mathbf{g}_{1}\left( \mathbf{\hat{v}}_{F},\mathbf{r},\varepsilon
_{m}\right) \right) \right\rangle  \label{spin-current}
\end{equation}
where, $\mathbf{{\hat{e}}}_{i}\mathbf{=}\left( \hat{\mathbf{x}},\hat{\mathbf{%
y}},\hat{\mathbf{z}}\right).$ We assume that the order parameter
does not depend on the coordinates and in each half-space it
equals its value (\ref {Bulk order parameter}) far from the
interface in the left or right bulks.
\begin{figure}[tbp]
\includegraphics[width=0.5\columnwidth]{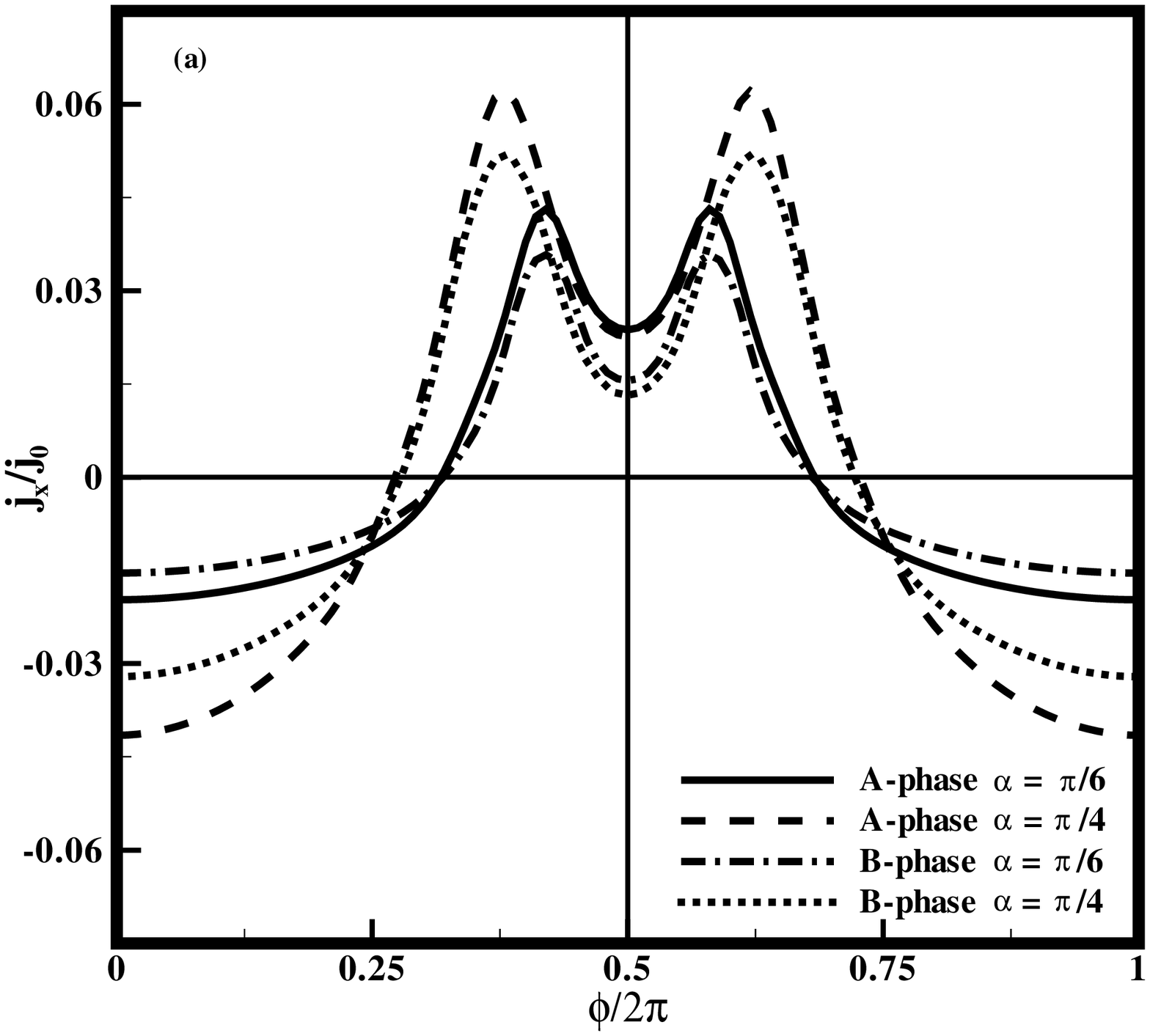}
\includegraphics[width=0.5\columnwidth]{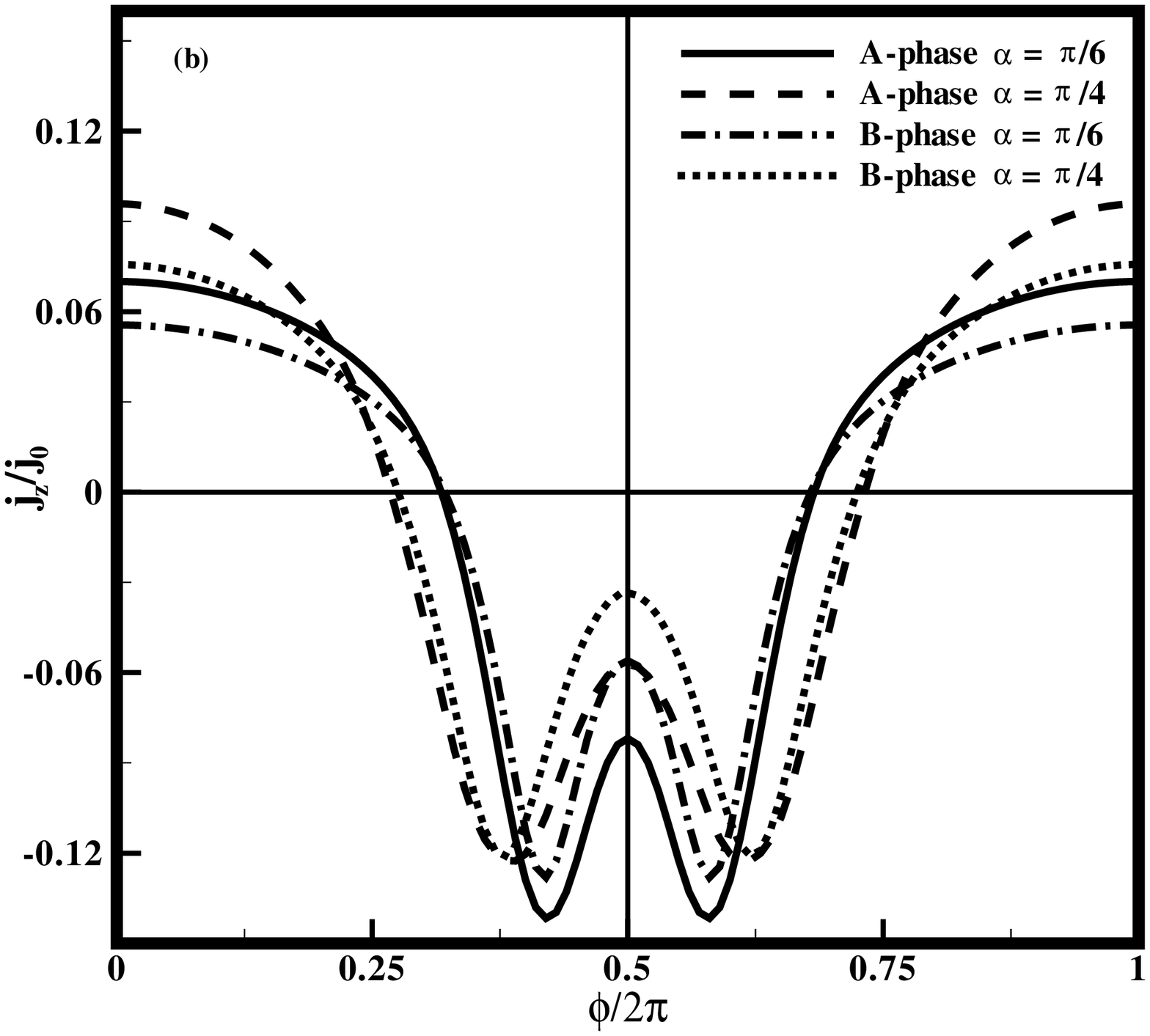}
\caption{Components of charge current tangential to the interface
versus the phase difference $\protect\phi $ for $A$ and
$B-$phases, geometry (ii), $T/T_{c}=0.08$ and the different
misorientations. Part (a) is plotted for $x-$component and part
(b) for $z-$component.} \label{fig3}
\end{figure}
For such a model, the current-phase dependence of a Josephson
junction can be calculated analytically. It enables us to analyze
the main features of current-phase dependence for different models
of the order parameter of ``$(p+h)-$wave'' superconductivity.
The Eilenberger equations (\ref{Eilenberger}) for Green functions $\breve{g}$%
, which are supplemented by the condition of continuity of
solutions across the interface, $y=0$, and the boundary conditions
at the bulks, are solved for a non-self-consistent model of the
order parameter analytically. Two diagonal terms of the Green
matrix which determine the current densities at the interface,
$y=0$, are shown below. For the term relating to the charge
current we obtain:
\begin{equation}
g_{1}\left( 0\right) =\frac{\varepsilon _{m}(\Omega _{1}+\Omega
_{2})\cos \beta +i\eta (\Omega _{1}\Omega _{2}+\varepsilon
_{m}^{2})\sin \beta }{i\eta \varepsilon _{m}(\Omega _{1}+\Omega
_{2})\sin \beta +(\Omega _{1}\Omega
_{2}+\varepsilon _{m}^{2})\cos \beta +\mathbf{\Delta }_{1}\mathbf{\Delta }%
_{2}},  \label{charge-term}
\end{equation}
and for the case of spin current we have:
$$\mathbf{g_{1}}\left(0\right)=\mathbf{\Delta
}_{1}\times \mathbf{\Delta}_{2}$$
\begin{equation}
\hspace{-2.5cm}\frac{(B-1)^{2}(\eta \Omega
_{1}+\varepsilon_{m})(\eta \Omega _{2}+\varepsilon _{m})\exp
(i\beta )-(B+1)^{2}(\eta \Omega _{2}-\varepsilon _{m})(\eta \Omega
_{1}-\varepsilon _{m})\exp (-i\beta
)}{{2\eta(A+B)\left|\mathbf{\Delta }_{1}\right|
^{2}\left|\mathbf{\Delta }_{2}\right| ^{2}}}
\end{equation}\label{spin-term}
where $\eta =sgn\left( v_{y}\right) $, $\Omega
_{n}=\sqrt{\varepsilon _{m}^{2}+\left| \mathbf{\Delta }_{n}\right|
^{2}}$, $\beta =\psi _{1}-\psi _{2}+\phi $,
\begin{equation}
B=\frac{\eta \varepsilon _{m}(\Omega _{1}+\Omega _{2})\cos \beta
+i(\Omega _{1}\Omega _{2}+\varepsilon _{m}^{2})\sin \beta }{i\eta
\varepsilon _{m}(\Omega _{1}+\Omega _{2})\sin \beta +(\Omega
_{1}\Omega _{2}+\varepsilon _{m}^{2})\cos \beta +\mathbf{\Delta
}_{1}\mathbf{\Delta }_{2}} \label{mathematics-term}
\end{equation}
and
\begin{equation}
A=\frac{\mathbf{\Delta }_{1}\mathbf{\Delta
}_{2}}{2}\left[\frac{(B-1)\exp (i\beta )}{(\eta \Omega
_{1}-\varepsilon _{m})(\eta \Omega _{2}-\varepsilon
_{m})}+\frac{(B+1)\exp (-i\beta )}{(\eta \Omega _{1}+\varepsilon
_{m})(\eta \Omega _{2}+\varepsilon _{m})}\right].
\end{equation}
Also, $n=1,2,$ label the left and right half-spaces respectively.
We consider a rotation $\breve{R}$ only in the right
superconductor (see, Fig.\ref{fig1}), i.e.,
$\mathbf{d}_{2}(\hat{\mathbf{k}})=\breve{R}\mathbf{d}_{1}(\breve{R}^{-1}\hat{%
\mathbf{k}});$ $\hat{\mathbf{k}}$ is the unit vector in the
momentum space. The crystallographic $c$-axis in the left
half-space is selected parallel to the partition between the
superconductors (along the $z$-axis in
Figs.\ref{phases},\ref{fig1}). To
illustrate the results obtained by computing the formula (\ref{charge-term}%
), we plot the current-phase diagrams for different models of the ``$%
(p+h)-$wave'' pairing symmetry (\ref{A-phase},\ref{B-phase}) and
for two different geometries. These geometries correspond to the
different orientations of the crystals on the right and left sides
of the interface (Fig.\ref{fig1}):\newline
(i) The basal $ab$-plane on the right side has been rotated around the $c$%
-axis by $\alpha $; $\hat{\mathbf{c}}_{1}\Vert
\hat{\mathbf{c}}_{2}$.\newline (ii) The $c$-axis on the right side
is rotated around the $b$-axis by $
\alpha $ ($y$-axis in Fig.\ref{fig1}); $\hat{\mathbf{b}}_{1}\Vert \hat{\mathbf{b}}%
_{2} $.\newline Further calculations require a certain model of
gap vector (order parameter vector) $\mathbf{d}$.
\section{Analysis of numerical results}
\label{section4} In the present paper, two forms of
``$(p+h)-$wave'' unitary gap vector $\mathbf{d}$ in
$PrOs_{4}Sb_{12}$ are considered. The first model to explain the
properties of the $A$-phase of $PrOs_{4}Sb_{12}$ is (left side of
Fig.\ref{phases}):
\begin{equation}
\mathbf{d}=\Delta _{0}(T)(k_{x}+ik_{y})\frac{3}{2}(1-\hat{k}_{x}^{4}-\hat{k}%
_{y}^{4}-\hat{k}_{z}^{4})\hat{\mathbf{z}}.  \label{A-phase}
\end{equation}
The coordinate axes
$\hat{\mathbf{x}},\hat{\mathbf{y}},\hat{\mathbf{z}}$ are chosen along the
crystallographic axes $\hat{\mathbf{a}},%
\hat{\mathbf{b}},\hat{\mathbf{c}}$ on the left side of Fig.\ref{fig1}; $%
\widehat{\mathbf{k}}$ is the unit vector along $\mathbf{v}_{F}$.
The scalar function $\Delta _{0}=$ $\Delta _{0}\left( T\right)$ describes the dependence of the gap vector $%
\mathbf{d}$ on the temperature $T$. The second model to describe
the gap vector of the $B$-phase of $PrOs_{4}Sb_{12}$ is (right
side of Fig.\ref{phases}):
\begin{equation}
\mathbf{d}=\Delta
_{0}(T)(k_{x}+ik_{y})(1-\hat{k}_{z}^{4})\hat{\mathbf{z}}.
\label{B-phase}
\end{equation}
\begin{figure}[tbp]
\includegraphics[width=0.5\columnwidth]{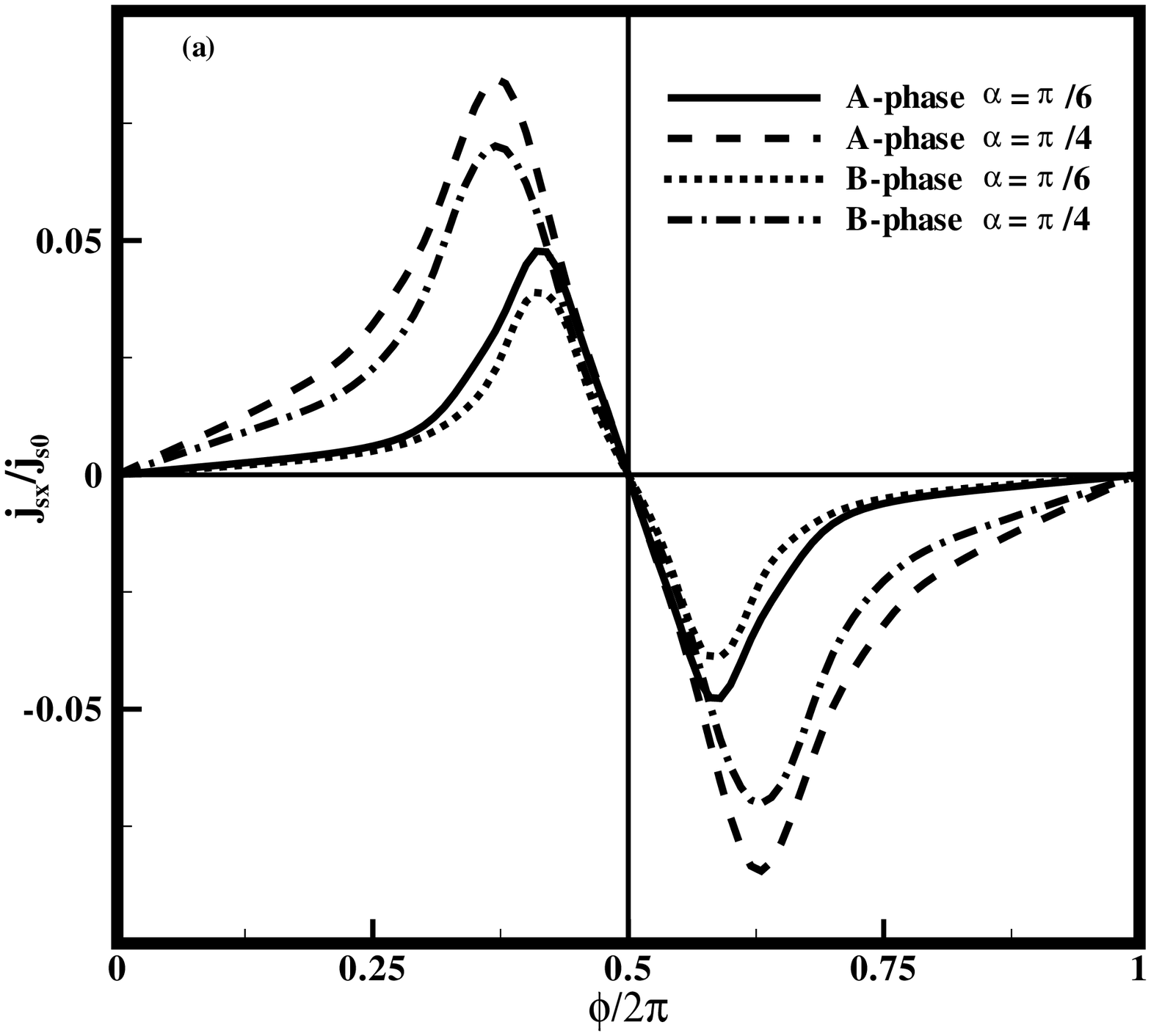}
\includegraphics[width=0.5\columnwidth]{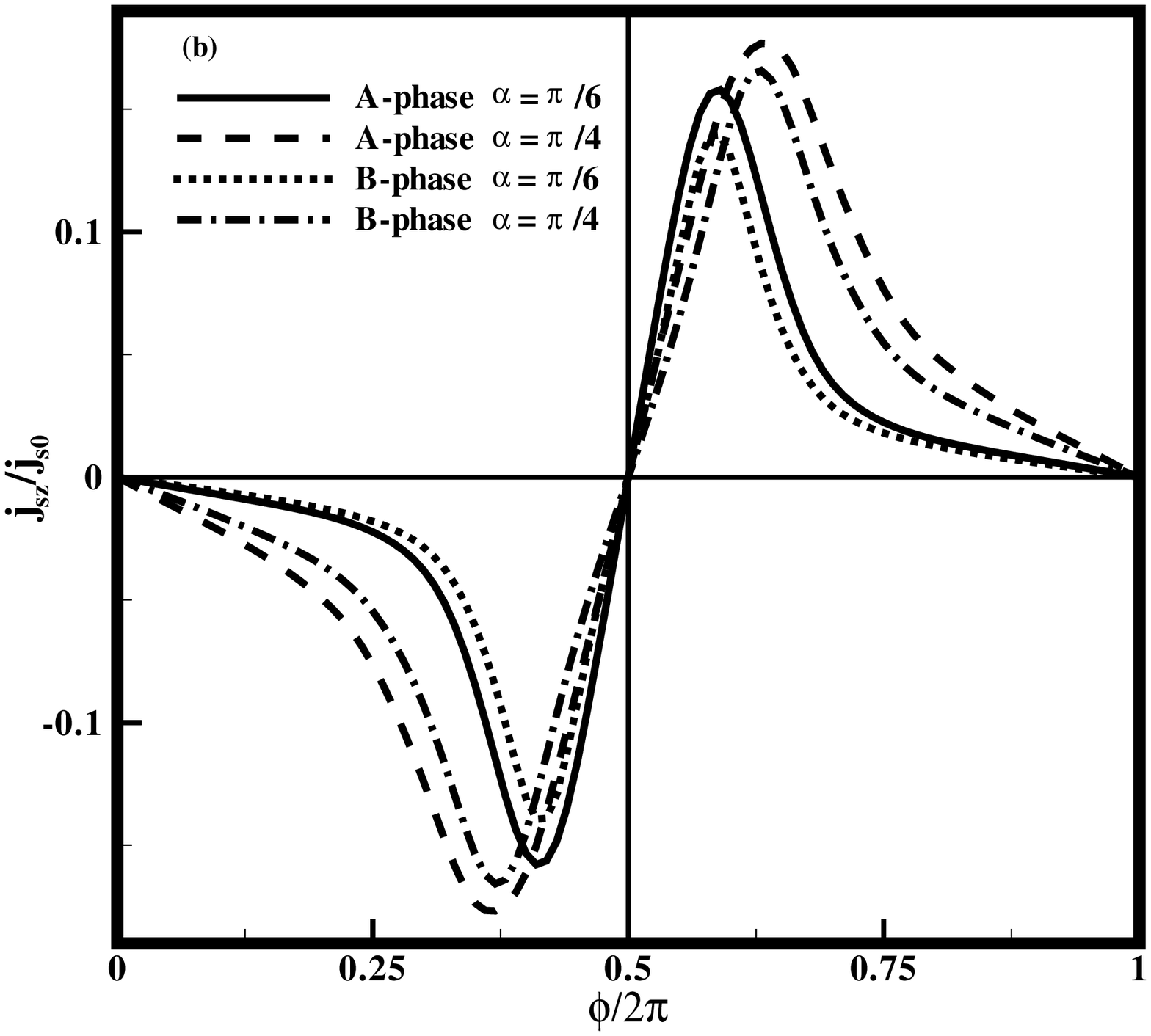}
\caption{Components of spin current ($s_{y}$) tangential to the
interface versus the phase difference $\protect\phi $ for geometry
(ii), $\frac{T}{T_{c}}=0.08$ and different misorientations between
the $A$ and $B-$phase of ``$(p+h)-$wave.'' Part (a) is plotted for
$x-$component
 and part (b) for $z-$component. Spin currents are given in units of
$j_{s0}=\frac{\protect\pi }{4}\hbar N(0)v_{F}\Delta _{0}(0).$}
\label{fig4}
\end{figure}
Our numerical calculations are done at the low temperatures,
$T/T_{c}=0.08$, and we have used the formulas $\ln (\Delta
(T)/\Delta (0))=-\frac{7\pi \zeta (3)}{8}(\frac{T}{\Delta
(0)})^{3}$ for the $A-$phase and $\ln (\Delta (T)/\Delta
(0))=-\frac{135\pi \zeta (3)}{512}(\frac{T}{\Delta (0)})^{3}$ for
the $B-$phase, from the paper \cite{Maki3}, for temperature
dependence of the gap functions $\Delta _{0}(T)$ at the low
temperatures ($T<<T_{c}$). Also, in the paper \cite{Maki3} the
value of $\Delta _{0}(T)$ has been calculated in terms of the
critical temperature for both $A$ and $B-$ phases. They are
$\Delta (0)/T_{c}=2.34$ and $\Delta (0)/T_{c}=1.93$ for $A$ and
$B-$ phases, respectively. Using these two models of order
parameters (\ref{A-phase}, \ref{B-phase}) and solution to the
Eilenberger equations (\ref{charge-term}, \ref{spin-term}), we
have calculated the current density at the interface numerically.
These numerical results are listed below:\newline 1) In part (a)
of Fig.\ref{fig2}, the component of current normal to the
interface of
current, which is known as the Josephson current, is plotted for both $A$ and $%
B-$phases, geometry (i), misorientations $\alpha =\pi /4$ and
$\alpha =\pi /6 $. It is observed that the critical values of
current for the $B-$phase is larger than the $A-$phase. Also,
unlike Josephson junction between the conventional
superconductors, here, at $\phi =0$, the current is not zero. The
current is zero at the phase difference value $\phi =\phi _{0}$,
which depends on the misorientation between the gap vectors. In
Fig.\ref{fig2}, the value of the spontaneous phase difference
$\phi _{0}$ is close to misorientation $\alpha .$\newline
2) In part (b) of Fig.\ref{fig2}, the Josephson current is plotted for both $A$ and $B-$%
phases, geometry (ii) and different misorientations. Again, the
maximum value of the current for the $B-$phase is larger than for
$A-$phase. Increasing the misorientation between the gap vectors,
the maximum value of the current decreases. It is demonstrated that at the phase difference values $%
\phi =0$, $\phi =\pi $ and $\phi =2\pi $, the Josephson current is
zero while both spontaneous and spin currents are not zero and
have a finite value. Increasing the misorientation between the gap
vectors decreases the derivative ($\frac{dj_{y}}{d\phi }$) of the
current with respect to the phase difference  close to $\phi =\pi
$, which is known as the \textbf{SQUID} sensitivity and it is
important from the application point of view.\newline 3) In
Fig.\ref{fig3}, the tangential components of the charge current
($x$ and $z-$components) in terms of the phase difference $\phi $
are plotted.
\begin{figure}[tbp]
\includegraphics[width=0.5\columnwidth]{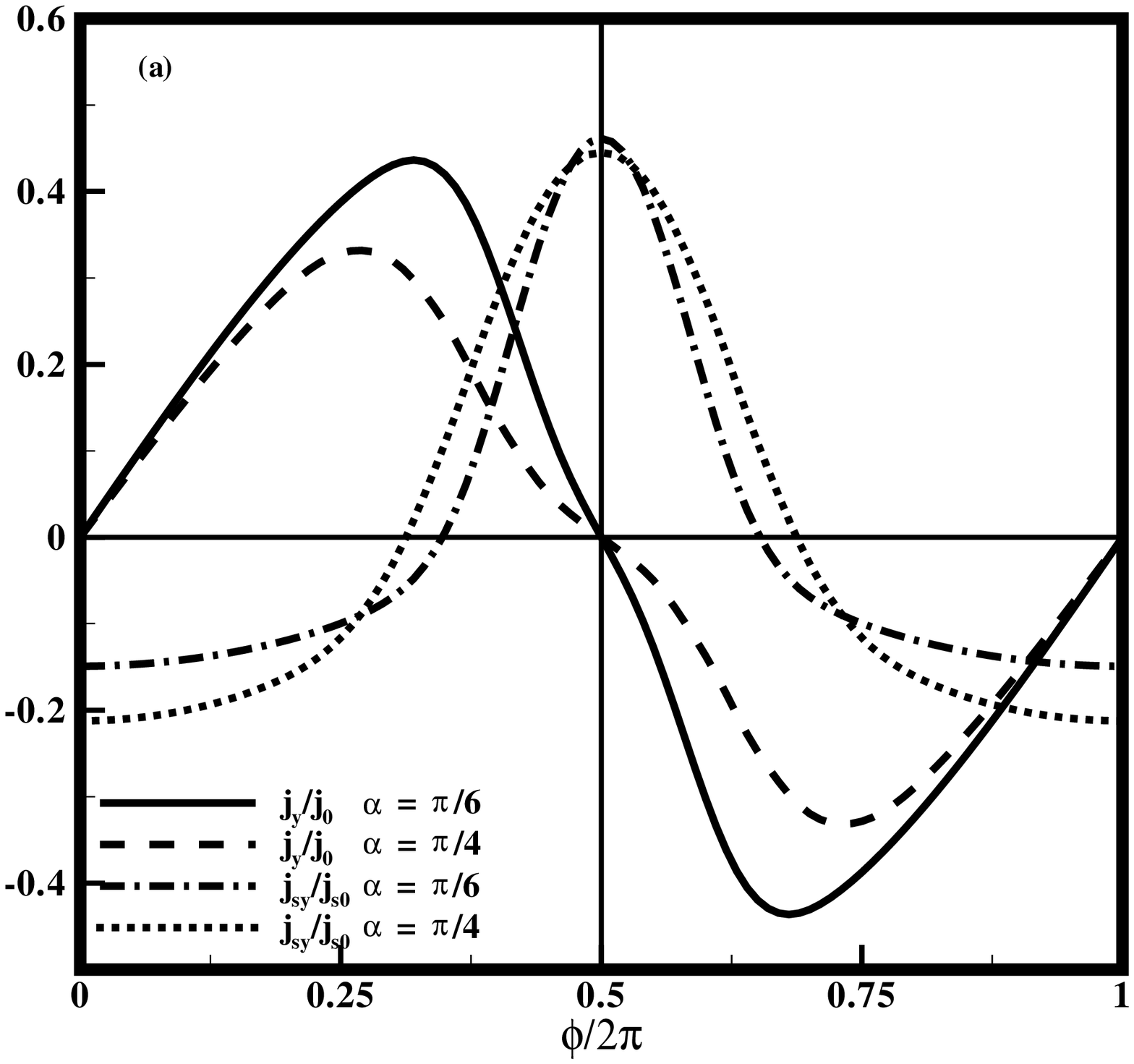}
\includegraphics[width=0.5\columnwidth]{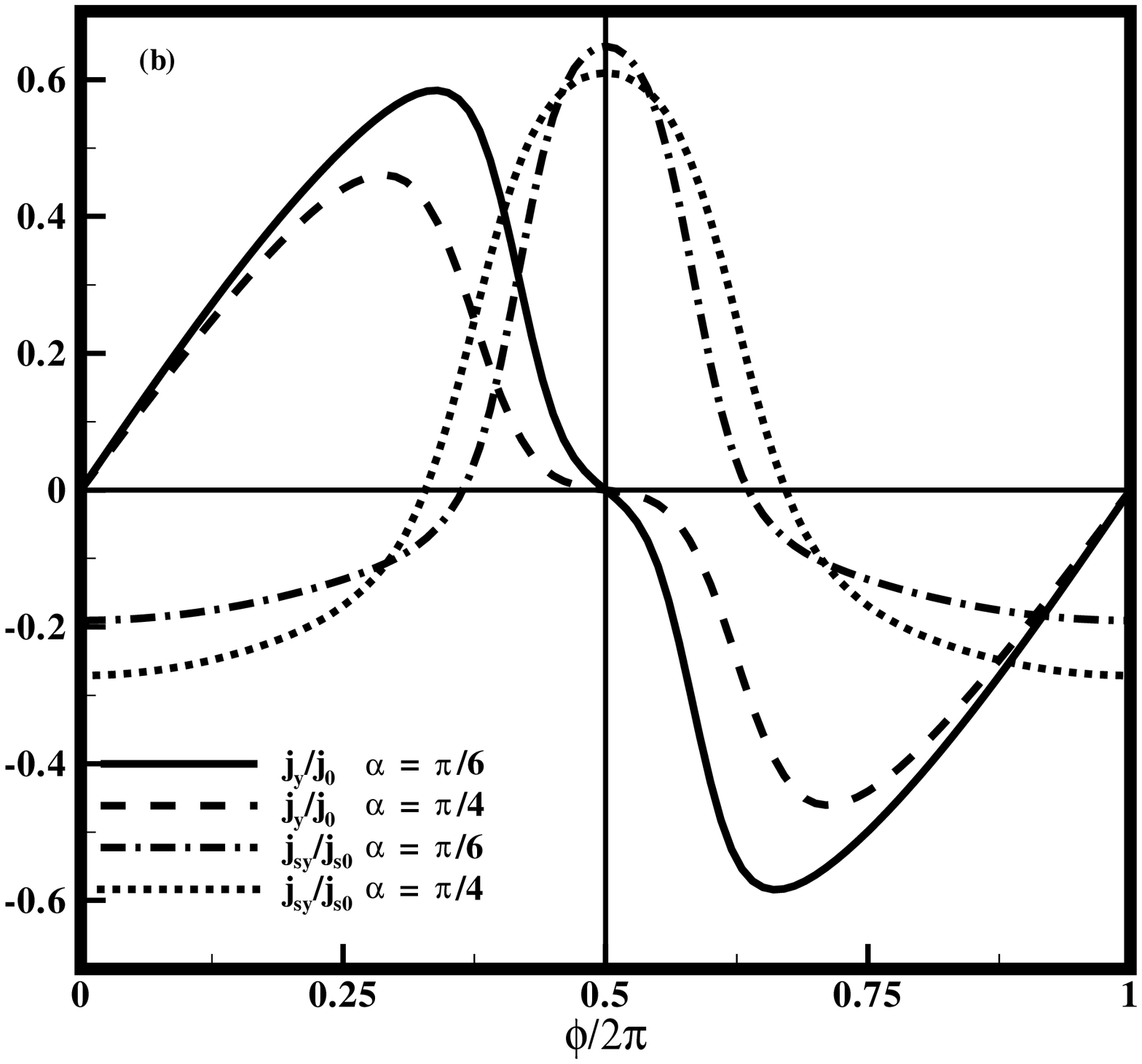}
\caption{Component of charge and spin currents ($j_{y}$ and $%
j_{sy}$) normal to the interface versus the phase difference
$\protect\phi $ for geometry (ii), $\frac{T}{T_{c}}=0.08$, $\protect\alpha =\frac{\protect\pi }{6}$ and $%
\protect\alpha =\frac{\protect\pi }{4}.$ Part (a) is plotted for
$A-$phase and part (b) for $B-$phase.} \label{fig5}
\end{figure}
It is seen that at $\phi =0$, $\phi =\pi $ and $\phi =2\pi ,$ at
which the Josephson current is zero, the parallel spontaneous
currents have finite values. Although the normal component
of charge current (see part (b) of Fig.\ref{fig2}) is an odd
function of the phase difference with respect to the line of $\phi
=\pi $ while the parallel charge currents for this geometry (ii)
are even functions of the phase difference with respect to $\phi
=\pi $ (compare part (b) of Fig.\ref{fig2} with
 Fig.\ref{fig3}).\newline
4) In Fig.\ref{fig4}, the tangential components of the spin
($s_{y}$) current are plotted in terms of the phase difference,
for geometry (ii) and different misorientations. By increasing the
misorientation the maximum value of the spin current increases. In
spite of the charge current for this state, the spin current at
the phase differences $\phi =0$, $\phi =\pi $ and $\phi =2\pi $ is
zero exactly(compare Fig.\ref{fig3} with Fig.\ref{fig4}).\newline
5) In Fig.\ref{fig5}, the normal component of charge and spin
current ($j_{y}$ and $j_{s_{y}}$) are plotted for different
misorientations and $A$ and $B-$phases respectively. An
interesting case in our observations, is the finite value of
normal spin current at $\phi =0$, $\phi =\pi $ and $\phi =2\pi $
at which the normal charge current ($j_{y}$) is zero (see
Fig.\ref{fig5}).
\newline
 6) In part (a) of Fig.\ref{fig6}, the
Josephson current is plotted in terms of the phase difference for the case of $p-$wave, and $A$ and $B-$phases of ``$(p+h)-$%
wave''. The $p-$wave pairing symmetry as the first candidate for
the superconducting state in Sr$_{2}$RuO$_{4}$ is as follows
\cite{Rice}:
\begin{equation}
\mathbf{d}=\Delta _{0}(T)(k_{x}+ik_{y})\hat{\mathbf{z}}
\label{p-wave}
\end{equation}
It is observed that the maximum value of the Josephson current
($j_{y}$) of the
junction between the $p-$wave superconductors, is larger than for the $B-$%
phase of ``$(p+h)-$wave'' and the Josephson current of $B-$phase
is larger than its value for the $A-$phase counterpart. Also, the
place of the zero of the charge current for these three types of
superconductor (geometry (i)) is the same. It is at the
spontaneous phase difference which is close to the misorientation
$\phi _{0}=\alpha $ (see, part (a) of Fig.\ref{fig2} and part (a)
of Fig.\ref{fig6}).\newline 7) In part (b) of Fig.\ref{fig6}, the
normal component of the spin current is plotted for $p-$wave, $A$
and $B-$phases of ``$(p+h)-$wave'' pairing
symmetries and for a specified value of misorientation $\alpha =\frac{\pi }{4%
}$. In both (a) and (b) parts of Fig.\ref{fig6}, the maximum value of
the current of junction between the $p-$wave superconductors is larger
than for the $B-$phase. For $B-$phase the current has the maximum
value, which is larger than the value for junction between the
``$(p+h)-$wave'' superconductors in the $A-$phase. This different
characteristic of the current-phase diagrams enables us to distinguish
between the three states.
Also, it is observed that at phase differences $\phi =0$,
$\phi =\pi $ and $\phi =2\pi $, the spin current has a finite
value which may have its maximum value. This is a counterpart of
part (b) of Fig.\ref{fig2}, in which the charge currents are zero at the
mentioned values of the phase difference, but the spin current has
a finite value.\newline

Furthermore, our analytical and numerical calculations have shown
that the origin of the spin current is misorientation between the
gap vectors [cross product in Eq.(\ref{spin-term})]. Because the
geometry (i) is a rotation by $ \alpha $ around the
$\mathbf{{\hat{z}}-}$axis and both of the left and right gap
vectors are in the same direction, the cross product between gap
vectors and consequently, the spin current for geometry (i) is
zero.
\begin{figure}[tbp]
\includegraphics[width=0.5\columnwidth]{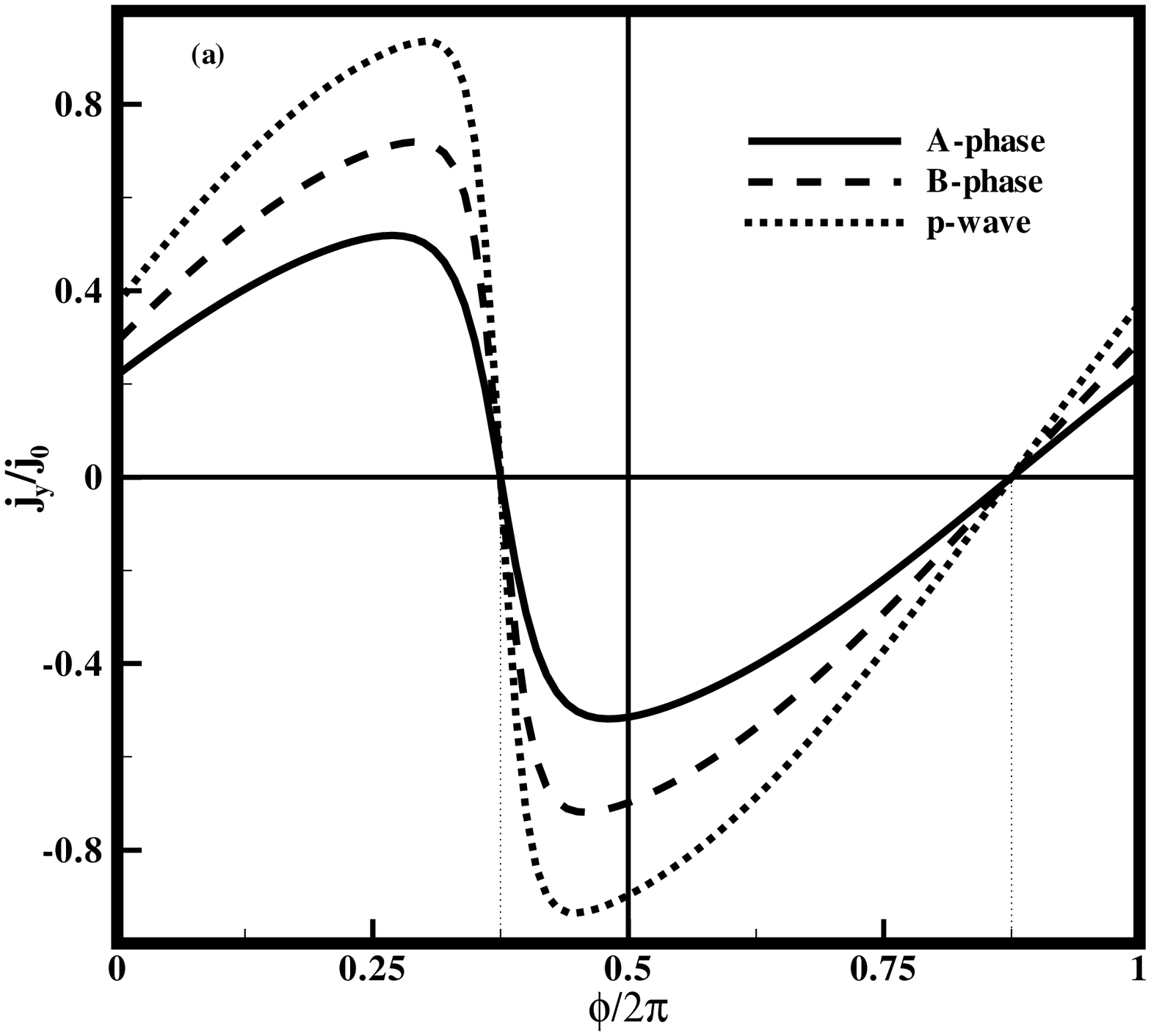}
\includegraphics[width=0.5\columnwidth]{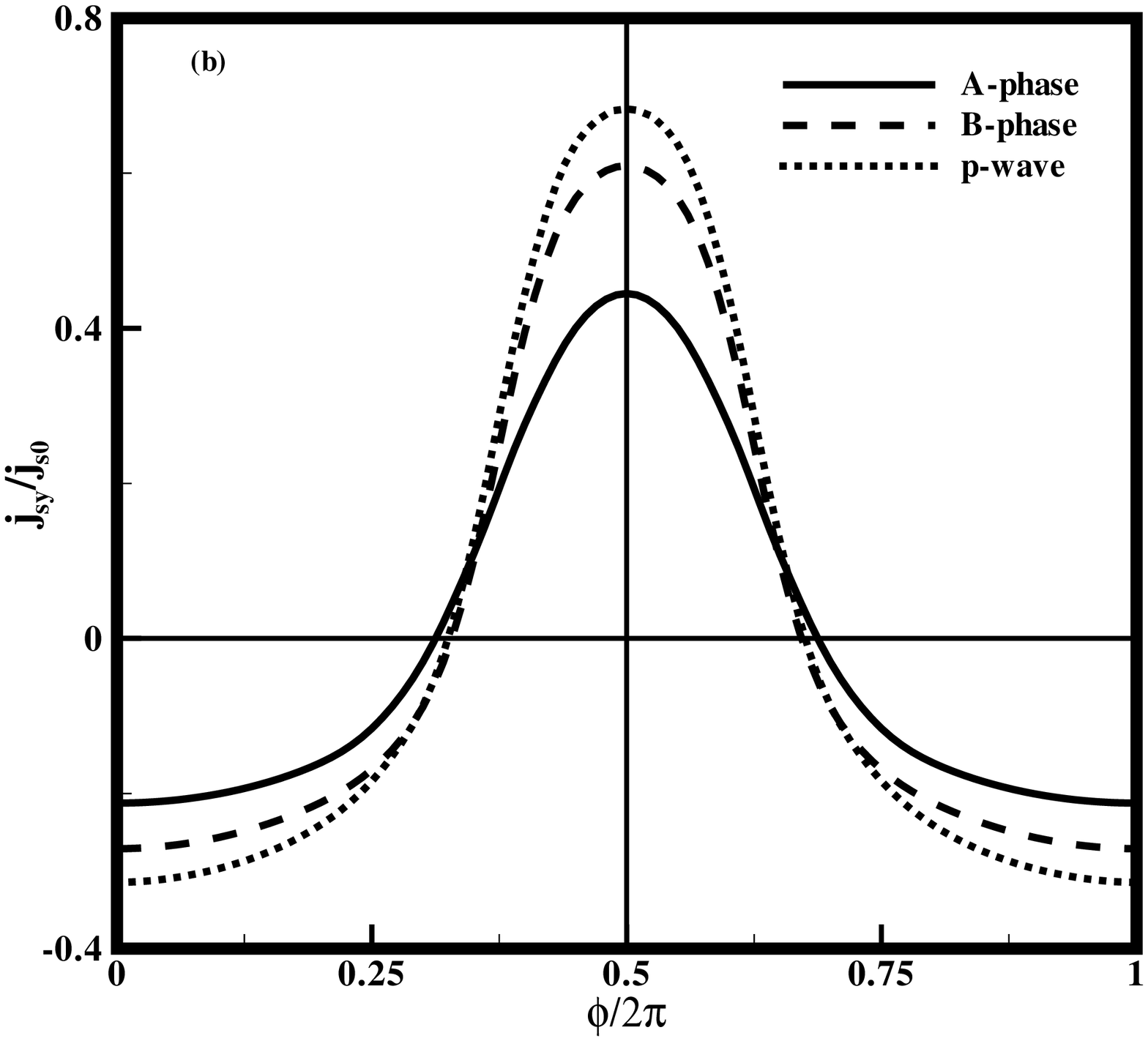}
\caption{Component of charge and spin currents normal to the
interface versus the phase difference $\protect\phi $,
$\frac{T}{T_{c}}=0.08$, $\protect\alpha =\frac{\protect\pi }{4}$,
$A-$phase, $B-$phase and $p-$wave pairing symmetry. Part (a) is
plotted for charge current and geometry (i) and part (b) is for
the case of spin current and geometry (ii)} \label{fig6}
\end{figure}
It is shown that in this structure ($y-$direction is normal
to the interface, $c-$axis is selected in the
$z-$direction and rotation is done around the $y-$direction) only
the current of the $s_{y}$ flows and other terms of the spin
current are absent. So, this kind of weak-link experiment can be
used as a filter for the polarization of spin transport. Since the
spin is a vector, the spin current is a tensor and we
have the current of spin $s_{y}$ in the three $\mathbf{{\hat{x}}}$, $\mathbf{%
{\hat{y}}}$ and $\mathbf{{\hat{z}}}$ directions.
\section{Conclusions}
\label{section5} We have theoretically studied the the spin and
charge transport in the ballistic Josephson junction in the model
of an ideal transparent interface between two misoriented
$PrOs_{4}Sb_{12}$ crystals with ``$(p+h)-$wave'' pairing symmetry,
which are subject to a phase difference $\phi $. Our analysis has
shown that the different misorientations and different models of
the gap vectors influence the spin and charge currents. This has
been shown for the charge current in the point contact between two
bulks of``$f-$wave'' superconductors in \cite{Mahmoodi} and for
the spin current in the weak link between ``$f$-wave''
superconductors in
 \cite{Rashedi1}. In this paper, it is shown that the
misorientation of the superconductors leads to a spontaneous phase
difference that corresponds to the zero Josephson current and to
the minimum of the weak link energy. This phase difference depends
on the misorientation angle. We have found a spontaneous charge
current tangential to the interface which is not equal to zero in
the absence of the Josephson current generally. It has been found
that the spin current is the result of the misorientation between
the gap vectors. Furthermore, it is observed that a certain model
of the gap vectors and geometries can be applied to polarize the
spin transport. Finally, as an interesting and new result, it is
observed that at certain values of the phase difference $\phi $,
the charge-current vanishes while the spin-current flows, although
the carriers of both spin and charge are the same (electrons). The
spatial variation of the phase of the order parameter plays a role
as the origin of the charge current and, similarly, due to the
broken $G^{spin-orbit}$ symmetry, a spatial difference of the gap
vectors in two half-spaces is the cause of spin currents. This is
because there is a position-dependent phase difference between
``spin up'' and ``spin down'' Cooper pairs and, although the total
charge current vanishes, there can be a net transfer of the spin.
Therefore, in our system, there is a discontinuous jump between
the gap vectors and, consequently the spin currents should
generally be present. For instance, if up-spin states and
down-spin states have a velocity in the opposite direction, the
charge currents cancel each other whereas the spin current is
being transported.
Mathematically speaking, $\mathbf{{j_{charge}}={j_{\uparrow }}+{j_{\downarrow }},{j_{spin}}=%
{j_{\uparrow }}-{j_{\downarrow }}}$, so it is possible to find the
state in which one of these current terms is zero and the other
term has a finite value \cite{Maekawa}. In conclusion, the spin
current in the absence of the charge current can be observed and
vice versa.

\end{document}